\documentclass[11pt]{article}
\usepackage{amsmath, amsthm, amssymb}    
\usepackage{bridges}			
\usepackage{graphicx}			
\usepackage[colorlinks=true, urlcolor=blue, citecolor=black, linkcolor=black]{hyperref}  

\usepackage{wrapfig}
\usepackage{pinlabel}
\usepackage{subfig}
\usepackage{color}

\newcommand{\TxI}{T^2\times\mathbb{I}}
\newcommand{\RR}{\mathbb{R}}

\title{Knotty knits are tangles on tori}

\author{Shashank G. Markande\textsuperscript{1} and Elisabetta A. Matsumoto\textsuperscript{2}
\vspace{10pt}\\
\textsuperscript{1}Georgia Institute of Technology; markande@gatech.edu\\
\textsuperscript{2}Georgia Institute of Technology; sabetta@gatech.edu} 

\date{}					

\begin{document}

\maketitle

\thispagestyle{empty}

\begin{abstract}

In this paper we outline a topological framework for constructing 2-periodic knitted stitches and an algebra for joining stitches together to form more complicated textiles. Our topological framework can be constructed from certain topological ``moves" which correspond to ``operations" that knitters make when they create a stitch. In knitting, unlike Jacquard weaves, a set of $n$ loops may be combined in topologically nontrivial ways to create $n$ stitches that are not pairwise associated. We define a \emph{swatch} as a construction that allows for these knitable knots.

\end{abstract}


\section*{Introduction}
\begin{wrapfigure}[12]{l}{0.55\textwidth}
\vspace{-25pt}
\centering
\subfloat[Schematic of a knitted fabric. It is a periodic structure of slip knots.]
 {\includegraphics[width=0.25\textwidth]{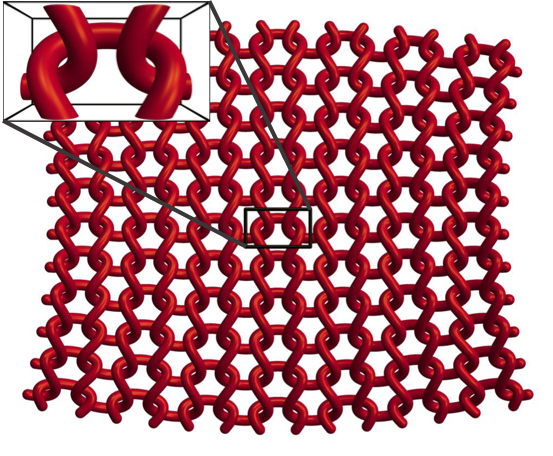} }
\subfloat[Textiles with intricate patterns are knit by combining slip knots in specific combinations.]
 {\includegraphics[width=0.25\textwidth]{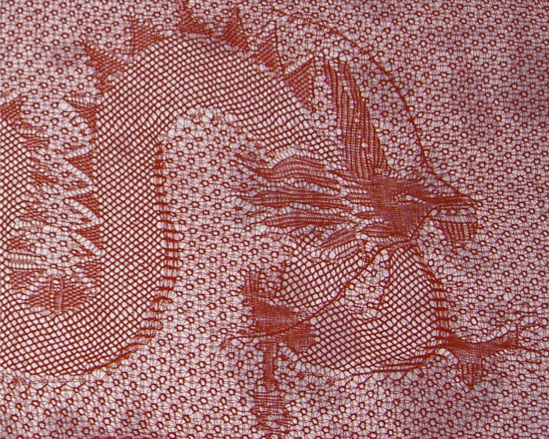} }

\vspace{-8pt}
\caption{} 
 \label{fig:knit_intro}
\vspace{-5pt}
\end{wrapfigure}
Imagine a 1D curve: entwine it back and forth so that it fills a 2D manifold which covers an arbitrary 3D object -- this computationally intensive materials challenge is realized in the ancient technology known as knitting. This process for making functional 2D materials from 1D yarns dates back to prehistory, with the oldest known examples found in Egypt from the 11th century CE  \cite{Albaron:1993}. Knitted textiles are ubiquitous as they are easy and affordable to create, lightweight, portable, flexible and stretchy. As with many functional materials, the key to knitting's extraordinary properties lies in its microstructure. The entangled structure of knitted textiles allows them to increase their length by over 100\% whilst barely stretching the constituent yarn.

From socks to performance textiles, sportswear to wearable electronics, knits are a ubiquitous part of everyday life. The geometry and topology of the knitted microstructure is responsible for many of these properties, even more so than their constituent fibers. But first, what constitutes a knit? 

\section*{Knits and purls}

Knits are composed of a periodic lattice of interlocking slip knots, also known as \emph{slip stitches}. At the most basic level, there is only one manipulation that constitutes knitting -- pulling a loop of yarn through another loop, (see Figure \ref{fig:knit_intro}a). There are two basic ``stitches" produced by this manipulation: a \emph{knit} stitch pulls a loop from the back of the fabric toward the front, whilst a loop pulled from the front of the fabric towards the back is called a \emph{purl} stitch. These stitches are actually the same; when viewed from the back, a knit stitch is a purl stitch.
 Combining these two motifs, there exist thousands of patterns of stitches with immense complexity, each of which has different elastic behavior (see Figure \ref{fig:knit_intro}b).

\begin{figure}[h!]
\centering
\subfloat[Knitting begins with loops on two needles. First you insert the right needle tip into the first loop on the left needle.]
 {\includegraphics[width=0.25\textwidth]{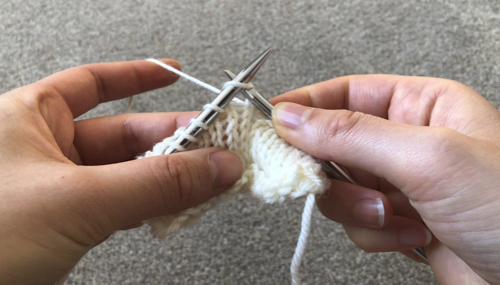} }
\subfloat[Then you wrap the free yarn around the right needle clockwise.]
 {\includegraphics[width=0.25\textwidth]{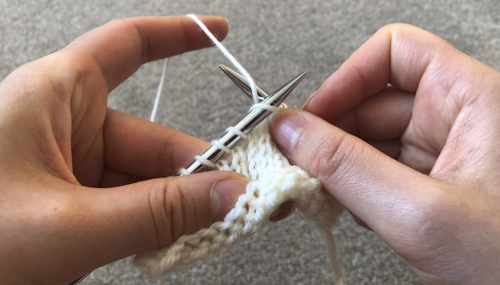} }
\subfloat[The newly formed loop of yarn gets pulled through the loop on the left needle.]
 {\includegraphics[width=0.25\textwidth]{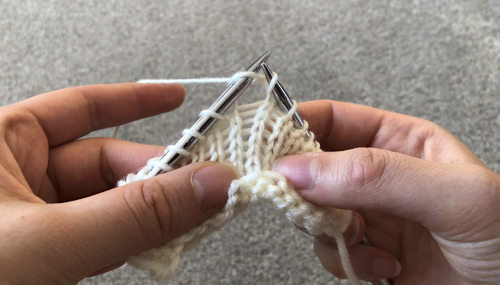} }
\subfloat[Lastly, you slide the loop off of the left needle. It is now captured by the loop you just made and both are caught on the right needle.]
 {\includegraphics[width=0.25\textwidth]{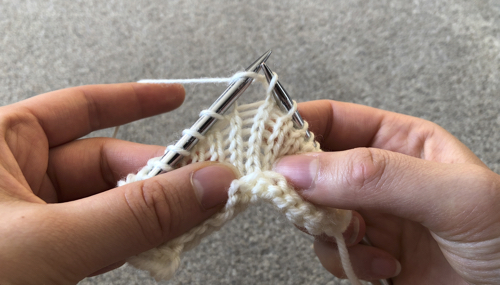} }

\vspace{-10pt}
 \caption{The process of hand knitting.}
\label{fig:knit_hands}
\end{figure}

 A piece of plain-knitted or weft-knit fabric contains only one thread which zigzags back and forth horizontally through
 the length of the fabric. The process of knitting threads slip stitches through loops from the previous row. Consecutive knitted stitches are connected to one another horizontally, a direction known as the \emph{course}. Knitted fabric is held together by a square lattice of these slip stitches -- rows are connected to each other vertically with slip stitches. Columns of slip stitches form along the vertical direction -- called the \emph{wale} -- connecting a single thread into a textile.

\begin{figure}[h!]
\vspace{-10pt}
\centering
\subfloat[{\rm Stockinette} fabric is formed by a lattice of knit stitches.]
 {\includegraphics[width=0.195\textwidth]{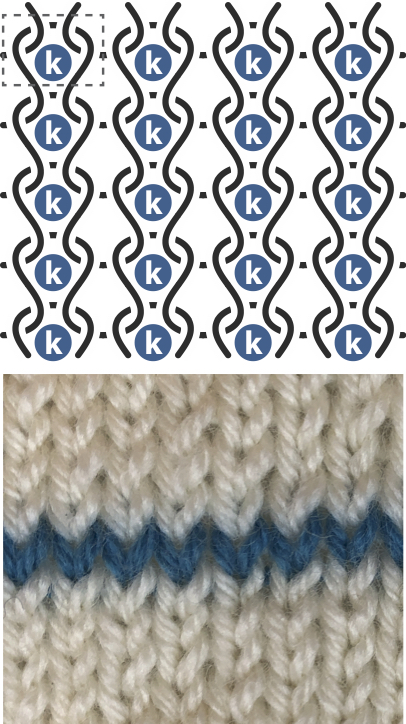} }
\subfloat[{\rm Reverse stockinette} fabric is formed by a lattice of purl stitches.]
 {\includegraphics[width=0.195\textwidth]{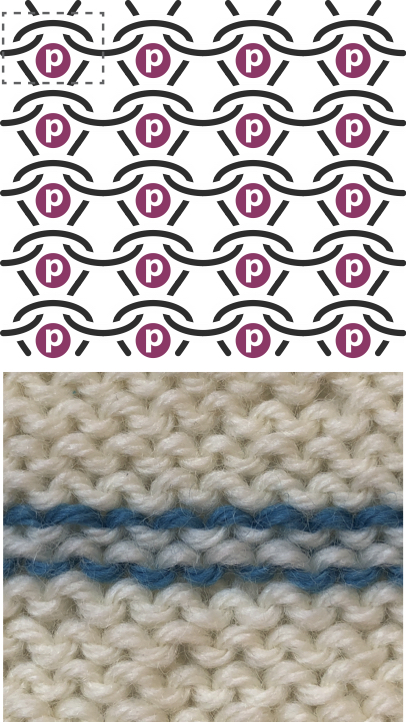} }
\subfloat[{\rm Garter} fabric alternates rows of all knit and all purl stitches.]
 {\includegraphics[width=0.195\textwidth]{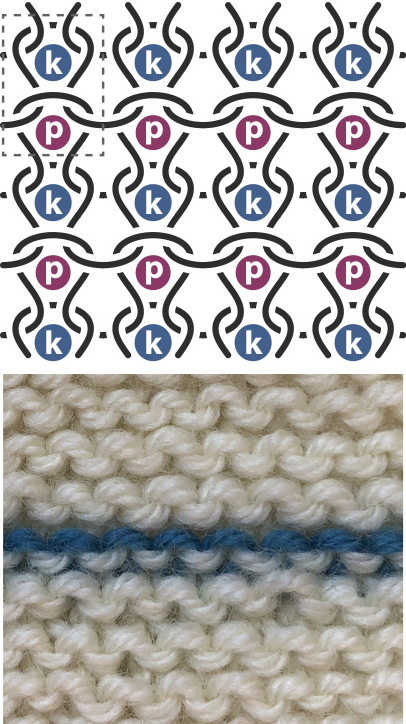} }
\subfloat[{\rm 1$\times$1 ribbing} alternates columns of all knits and all purls.]
 {\includegraphics[width=0.195\textwidth]{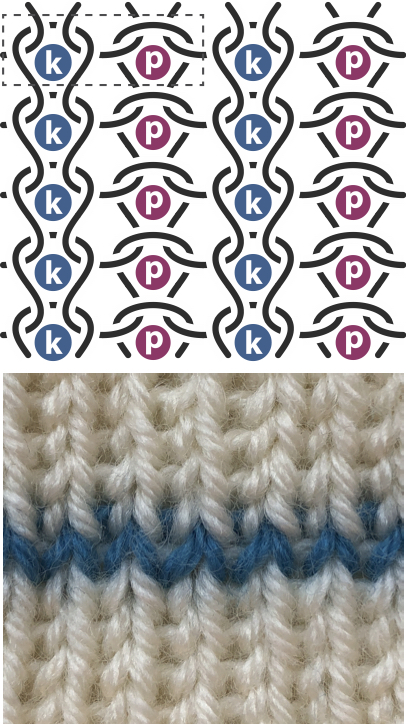} }
\subfloat[{\rm Seed} fabric is a checkerboard lattice of knits and purls.]
 {\includegraphics[width=0.195\textwidth]{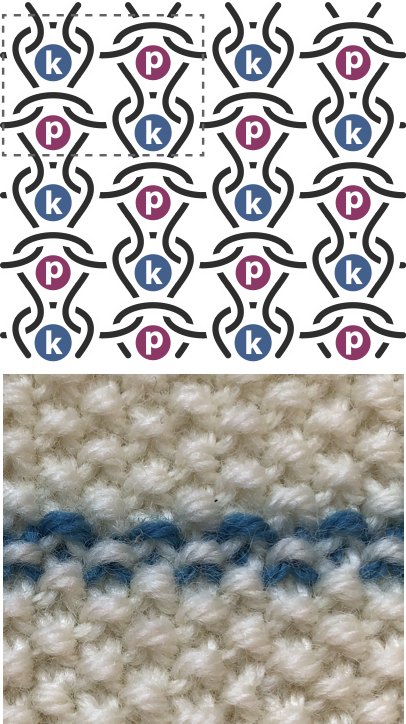} }

\vspace{-5pt}\caption{Common fabrics created using knit and purl stitches. Remarkably these fabrics all have very different elastic behaviors, despite being nearly topologically identical. (The exception to this is stockinette and reverse stockinette, which are related by rotational symmetry.)}
 \label{fig:knit_purl}
\end{figure}

Using solely knit and purl stitches, thousands of distinct fabrics can be created, each with different elastic properties. See Figure \ref{fig:knit_purl}. \emph{Stockinette} fabric\footnote{Sometimes called stockinette stitch. The term \emph{stitch} is used in two contexts in knitting parlance. \emph{Stitch} may refer to single stitch such as a knit or a purl, or it may refer to a fabric created by a small number of repeated stitches. The former definition shall be used throughout, and the term \emph{fabric} shall refer to a pattern of stitches.} 
 is created entirely of knit stitches (Figure \ref{fig:knit_purl}a). Likewise, \emph{reverse stockinette} is made from entirely purls (or by turning over stockinette fabric) (Figure \ref{fig:knit_purl}b). Stockinette and reverse stockinette have a preference for negative gaussian curvature. In both fabrics, the bottom and top curl towards the knit side of the fabric, whilst the left and right slides curl towards the purl side. \emph{$1\times1$ ribbing} alternates knits and purls keeping all stitches in each column the same (Figure \ref{fig:knit_purl}c). This fabric is very stretchy and has a corrugated appearance. Ribbing fabric is frequently used for cuffs and collars of garments. \emph{Garter} fabric alternates rows of knit stitches and purl stitches (Figure \ref{fig:knit_purl}d). \emph{Seed} fabric is a checkerboard lattice of knits and purl (Figure \ref{fig:knit_purl}e). The latter three fabrics lie flat because they have a rotational symmetry in the plane of the fabric that leaves the front and back of these fabrics indistinguishable. Stockinette and reverse stockinette fabrics lack this symmetry and the local deformation of each knit (or purl) stitch is compounded across the entire fabric, 
with the consequence that it curls. The local topology of stitches, as well as the order in which they appear in the fabric determines the local geometry of the fabric and, therefore, its elastic response.

\section*{Knits as knots in $\mathbb{T}^2\times I$}

\begin{figure}[h!]
\centering
\subfloat[The knitted stitch lives in the manifold $\mathbb{T}^2\times I$. Here, green sides are identified and red sides are identified.]
 {\includegraphics[width=0.245\textwidth]{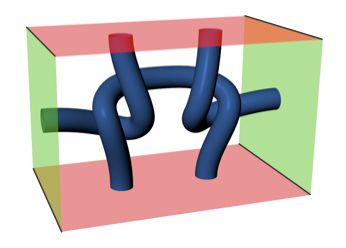} }
\subfloat[In order to see the knit stitch as a link in $S^3$, we created $\mathbb{T}^2\times I$ by subtracting the tubular neighborhood of a Hopf link from.]
 {\includegraphics[width=0.245\textwidth]{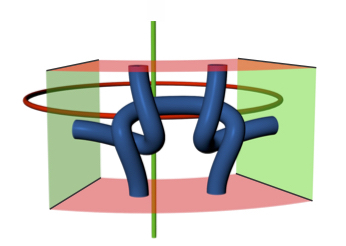} }
\subfloat[When the green faces are identified, the knit stitch must link with the green component of the Hopf link.]
 {\includegraphics[width=0.245\textwidth]{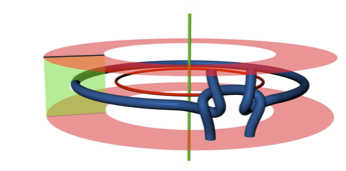} }
\subfloat[The green component of the Hopf link is truly an $S^1$ embedded in $S^3$.]
 {\includegraphics[width=0.245\textwidth]{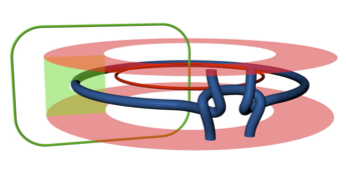} }\\
 
\vspace{-10pt} 
\subfloat[When the red faces are identified, the knit stitch must link with the red component of the Hopf link.]
 {\includegraphics[width=0.245\textwidth]{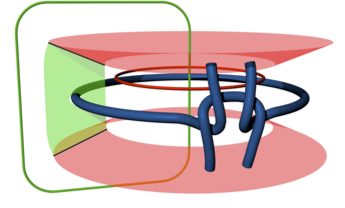} }
\subfloat[The green and red surfaces in $T^2 \times I$ are a pair of annuli that intersect along a single line.]
 {\includegraphics[width=0.245\textwidth]{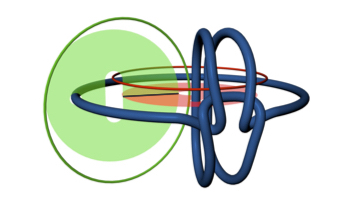} }
\subfloat[The 2-periodic knit stitch is now a three component link in $S^3$.]
 {\includegraphics[width=0.245\textwidth]{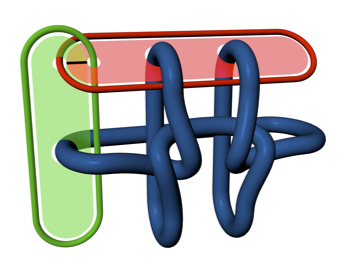} }
\subfloat[This planar projection shows the 2-periodic knit stitch in {\rm standard position}.]
 {\includegraphics[width=0.245\textwidth]{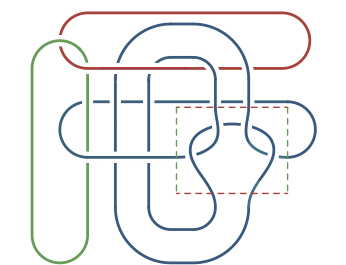} }

\vspace{-5pt}
\caption{2-periodic knit stitches naturally live in $T^2\times I$. However, we can construct 2-periodic knit stitches as three component links in $S^3$.}
 \label{fig:T2xI}
\vspace{-5pt}
\end{figure}

Topology and entanglement hold textiles together, yet knits are topologically trivial; because a knitted textile is comprised of slip knots, pulling a single loose thread can unravel the entire garment. Knitting is doubly periodic -- that is, it lives on a square lattice. Thus, invoking periodic boundary conditions leaves us with a knot that cannot be untangled, see Figure \ref{fig:knit_purl},\ref{fig:T2xI}a. To see this, note that it has nontrivial homology longitudinally, that is along the green direction in Figure \ref{fig:T2xI}a.

Knot theory provides us with a natural framework to study such entanglement problems. A \emph{knot} is a nontrivial embedding of a circle $S^1$ into $\mathbb{R}^3$. Likewise, a \emph{link} consists of two or more circles embedded in $\RR^3$. Two knots or links are topologically \emph{equivalent} if one can be transformed into the other via a deformation of the ambient space that does not involve cutting the knot or letting the string pass through itself. Knot theory studies topological descriptors of this equivalence. We seek to create an algebra for textile knots that can incorporate all possible types of slip-stitches compatible with knitting. This can handle finite samples and infinite fabrics made of repeated patterns of stitches. 

Knits, weaves and other 2-periodic textiles live naturally in a space homeomorphic to a thickened torus, $T^2\times \mathbb{I}$. We wish to study these textile knots in this natural space, thus we turn to  3-manifold topology. Any invariant of a the manifold created by removing a 
 tubular neighborhood $\mathcal{T}$ from around the knot $\mathcal{K}$ in the 3-sphere, denoted $S^3-\mathcal{T}_\mathcal{K},$ is also an invariant of the knot $\mathcal{K}$. When the knot is not embedded ambient euclidean space $\mathbb{R}^3$ (as is the case with textile knots living in $T^2 \times \mathbb{I}$), we can create the ambient manifold by removing a specific knot or link from $S^3$. In particular, $T^2\times \mathbb{I}$ is homeomorphic to $S^3$ minus a Hopf link, which is a pair of embedded circles which pass through each other's centers. 

Constructing our manifold as $S^3-\mathcal{T}_\mathcal{L},$ where $\mathcal{L}$ is the link composed of the textile knot $\mathcal{K}$ and the Hopf link, allows us to use the link editor in Snappy \cite{SnapPy} to create a triangulation of this 3-manifold. The following is a canonical construction of our manifold. We start with a knitted stitch in the thickened torus $\TxI$ (Figure \ref{fig:T2xI}a),
  where the pairs of green and pairs of red sides are identified. In Figure \ref{fig:T2xI}b, this is then put into $S^3-\mathcal{T}_{\textrm{(Hopf link)}}$, where the red and green tubes designate the Hopf link. Note, the green tube connects through infinity. The green sides of the thickened torus in Figure \ref{fig:T2xI}c connect by encircling the green circle of the Hopf link. This green cycle is resized to fit in the frame in Figure \ref{fig:T2xI}d. The final maneuver to connect up the knitted stitch, in Figure \ref{fig:T2xI}e,f, identifies the red faces with one another by wrapping around the red element of the Hopf link.

We now define \emph{standard position} for a link $\TxI$ which has been lifted into $S^3-\mathcal{T}_{\textrm{(Hopf link)}}$, see Figure \ref{fig:T2xI}g. Standard position is a canonical construction of the textile link in $S^3$. In standard position, the identified sides of the original thickened torus (red and green in Figure \ref{fig:T2xI}a) are now annuli. Each annulus has one boundary component isotopic to the component of the Hopf link of the corresponding color. The other boundary is punctured by the other component of the Hopf link. These annuli intersect one another along a curve that connects the two boundary components. The course direction punctures the green surface, and the wale punctures the red surface.

By converting this image into a two dimensional link diagram with planar crossings (Fig. \ref{fig:T2xI}h). In Fig. \ref{fig:T2xI}h, there is a dashed rectangle which corresponds to a flattened version of the original knot in $\TxI$. One might ask what conditions exist on knots in the dashed rectangle such that they are knitable? Hand knitters have an implicit notion of what a \emph{stitch} is -- a set of manipulations of existing loops and/or free yarn that ends when a loop is passed from the left needle to the right needle. Unfortunately, rigorizing this definition will always require a choice. Some ambient isotopies of a \emph{bight} -- a small continuous segment -- of yarn, might be too complex for a knitter to do using only two needles without additional equipment or scaffolding, however topologically, these would always be allowed. For example, twisting a stitch an arbitrarily large number of times or creating an arbitrarily long chain of single crochet are topologically consistent with being knitable.

For a knot to be knitable, it must be created from \emph{slip knots}, which are a class of ambient isotopies of a portion of the unit line with ends fixed created by pulling bights of that line through one another. In $\mathbb{T}^2\times I$, this class of knots has nontrivial homology around the longitude (shown in all diagrams here as the horizontal cycle) and trivial homology around the meridian (the vertical cycle, here). This implies that in a knitted textile, each row of stitches is connected together along one piece of yarn while neighboring rows are pairwise trivially linked. This is apparent in standard position. The knitted component of the link (blue) is pairwise linked with the green component of the Hopf link (the longitude) and is pairwise unlinked with the red component (the meridian), as shown in Figure \ref{fig:T2xI}h.

\begin{wrapfigure}[9]{r}{0.6\textwidth}
\vspace{-25pt}\subfloat[Ribbon knot.]{\includegraphics[height=0.18\textwidth]{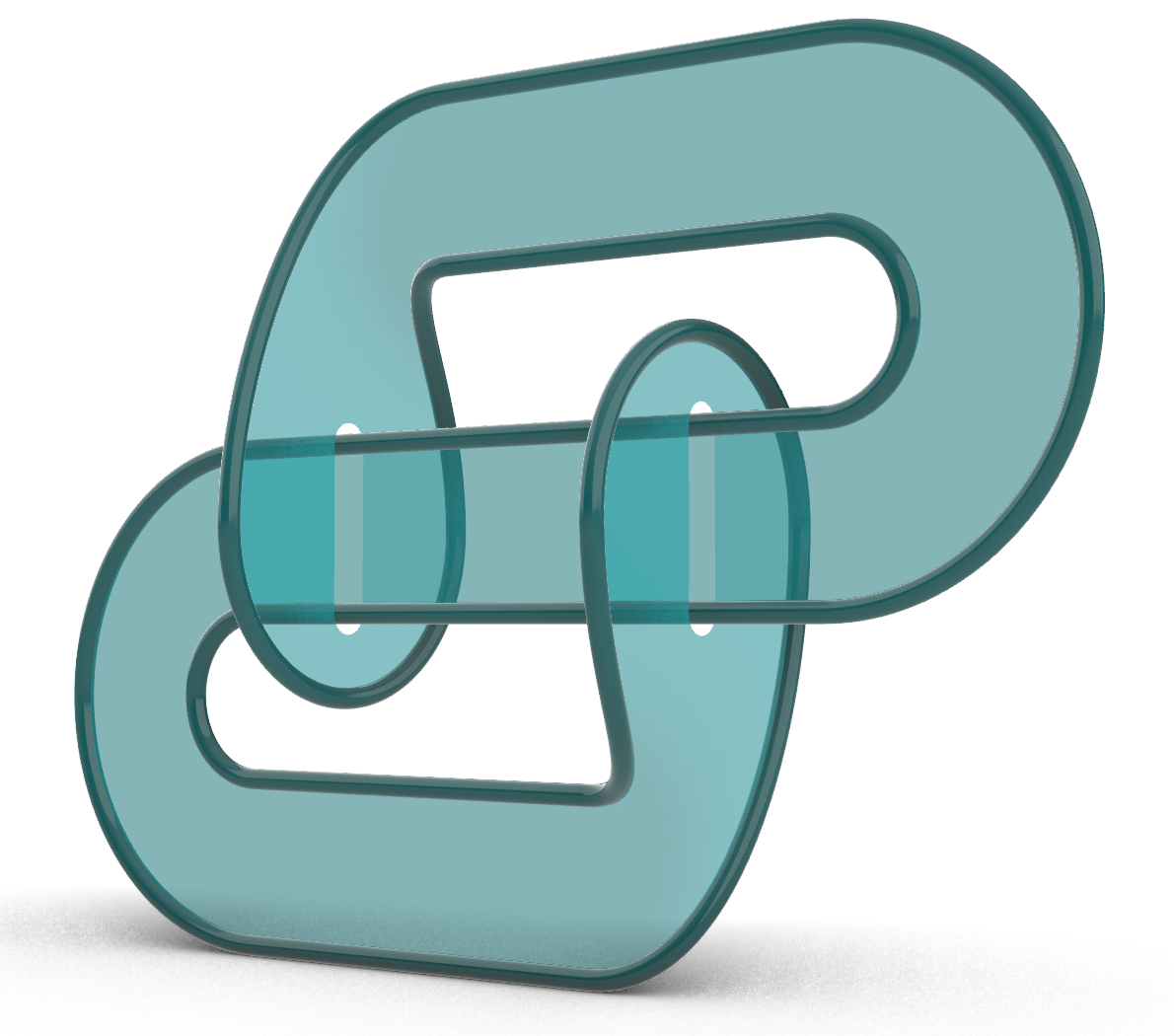} }
\subfloat[Cow-hitch schematic]{\includegraphics[height=0.18\textwidth]{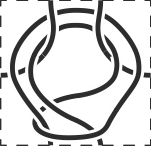} }
\hspace{5pt}
\subfloat[Cow-hitch knitted]{\includegraphics[height=0.18\textwidth]{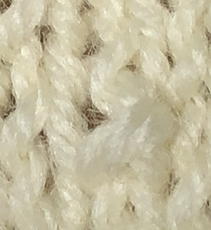} }
\caption{} 
 \label{fig:cow}
 \end{wrapfigure}
 An examination of many commonly used knitable stitches reveals that all share the property that they are \emph{ribbon}. \emph{Ribbon knots} are knots that bound a self-intersecting disk where all self intersections are \emph{ribbon singularities} -- places where the ribbon self intersects form curves that exist only in the interior of the spanning disk. Intuitively, this is not surprising, since all knits are formed by sliding bights of yarn through each other. We conjecture that all knits are ribbon. We will show later that being ribbon is a necessary, but not sufficient, condition for a knot to be knitable.

What types of ribbon knots can be turned into knitable stitches? A class of potentially knitable ribbon knots come from tying other knots or links into a bight and the knitting that into the next row. One example of such a stitch we call the \emph{cow-hitch} (shown in Figure \ref{fig:cow}). This stitch is made by tying a half hitch into the bight and then knitting through it.

\section*{Combining stitches using annulus sums}

\begin{figure}[h!]
\centering
\subfloat[A disjoint link in $S^3$.]
 {\includegraphics[width=0.25\textwidth]{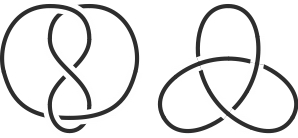} }\hspace{10pt}
 \subfloat[The two components of the link are joined by a band.]
 {\includegraphics[width=0.25\textwidth]{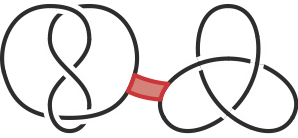} }\hspace{10pt}
 \subfloat[Band surgery swaps arcs long the edges of the band.]
 {\includegraphics[width=0.25\textwidth]{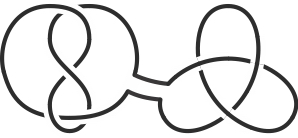} }\\
 
\vspace{-5pt}\caption{By joining 2-periodic knit stitches together in different ways we can generate the different fabrics in Figure \ref{fig:knit_purl}.}
 \label{fig:connect_sum}
\end{figure}

 Now that we have constructed a standard position for textile knots in $S^3$, we need to construct an algebra for adding different stitch types together to create fabrics, as in Figure \ref{fig:knit_purl}. In $S^3$, a \emph{connected sum} of two disjoint knots $\mathcal{K}_1$ and $\mathcal{K}_2$, denoted by $\mathcal{K}_1\#\mathcal{K}_2$, joins $\mathcal{K}_1$ and $\mathcal{K}_2$ according to the following procedure: (1) take planar projections of two knots (Figure \ref{fig:connect_sum}a), (2) find a rectangular patch where one pair of sides are arcs on each knot (Figure \ref{fig:connect_sum}b) and (3) join the knots by deleting the two sides of the knot in the rectangle and connecting the other pair of sides (Figure \ref{fig:connect_sum}c).\footnote{Note the general procedure of changing the connectivity of a knot or link according the a rectangle (as in steps (2) and (3)) is called \emph{band surgery}. This has many consequences for topological invariants.}
 For instance, the Alexander polynomial $V$ for $\mathcal{K}_1\#\mathcal{K}_2$, $V_{\mathcal{K}_1\#\mathcal{K}_2}=V_{\mathcal{K}_1}V_{\mathcal{K}_2}$ is a product of the Alexander polynomials for each individual knot, $V_{\mathcal{K}_1}$ and $V_{\mathcal{K}_2}$. This creates an algebra for building complexity of knots in $S^3$.
 
\begin{figure}[h!]
\vspace{-10pt}
\subfloat[Two 2-periodic knit stitches are joined in the $\mathbb{T}^2\times I$ model.]
 {\includegraphics[width=0.495\textwidth]{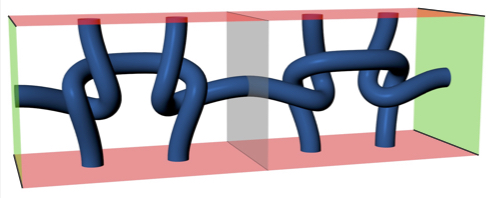} }
\subfloat[The same two stitches now joined in $S^3-\mathcal{T}_{\textrm{(Hopf link)}}$.]
 {\includegraphics[width=0.425\textwidth]{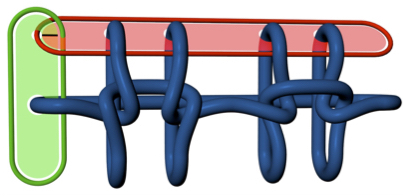} }

\caption{Two stitches joined horizontally to create $1\times1$ ribbing.}
\label{fig:horiz_join}
\vspace{-5pt}
\end{figure}
\begin{figure}[t!]
\centering
\subfloat[Adding stitches horizontally by cutting two tori along their meridians.
 ]
 {\begin{minipage}{0.245\textwidth}\vspace{-100pt}\includegraphics[width=1.\textwidth]{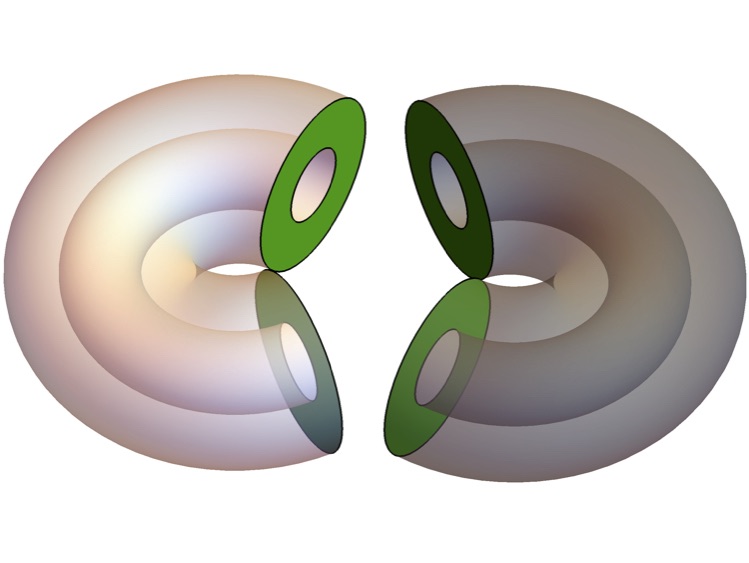}\end{minipage} }
\subfloat[The 3-manifolds are then glued together along the boundary annuli.]
 {\includegraphics[width=0.245\textwidth]{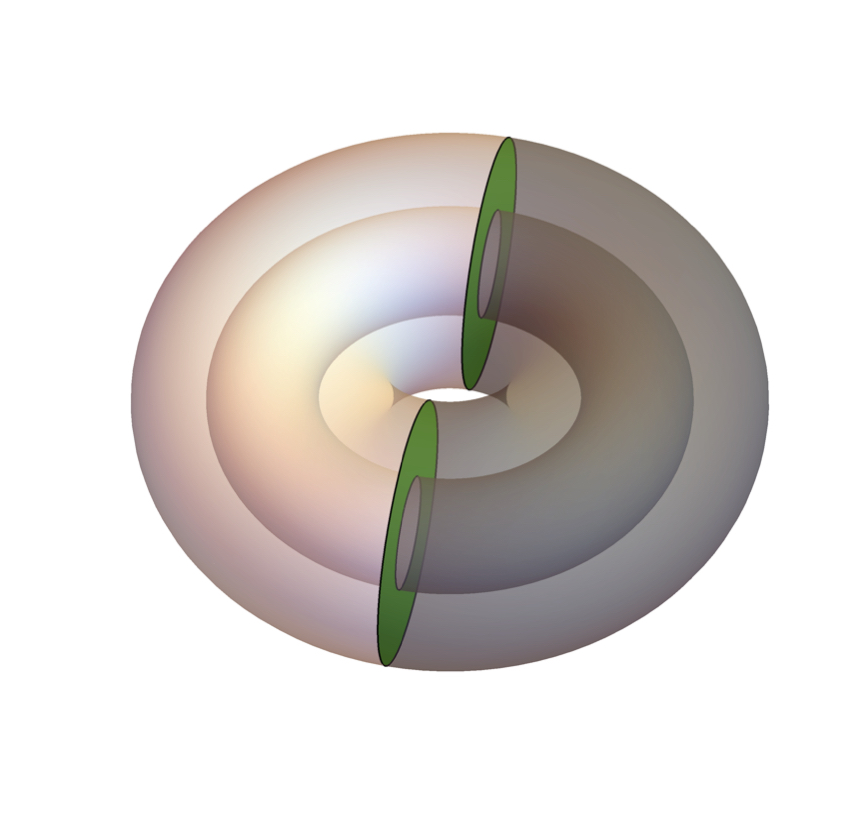} }
\subfloat[In the $S^3$ picture, this can be algebraically realized using band surgery.]
 {\begin{minipage}{.245\textwidth}\vspace{-90pt}\centering\includegraphics[width=0.95\textwidth]{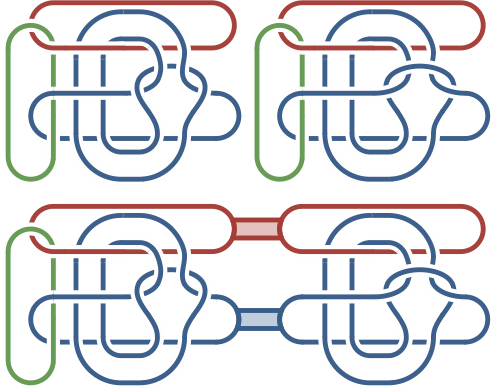}\end{minipage} }
 \subfloat[$1\times1$ ribbing from a meridional annulus sum.]
  {\begin{minipage}{0.245\textwidth}\vspace{-90pt}\centering\includegraphics[width=0.9\textwidth]{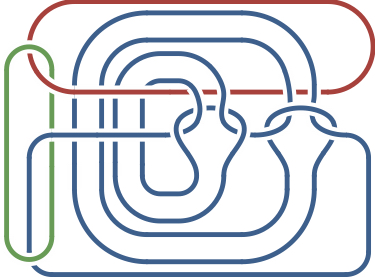} \end{minipage}}
\\

\vspace{90pt} 
\subfloat[Adding stitches vertically by cutting two tori along their longitudes.
]
 {\begin{minipage}{.245\textwidth}\vspace{-90pt}\centering\includegraphics[width=0.85\textwidth]{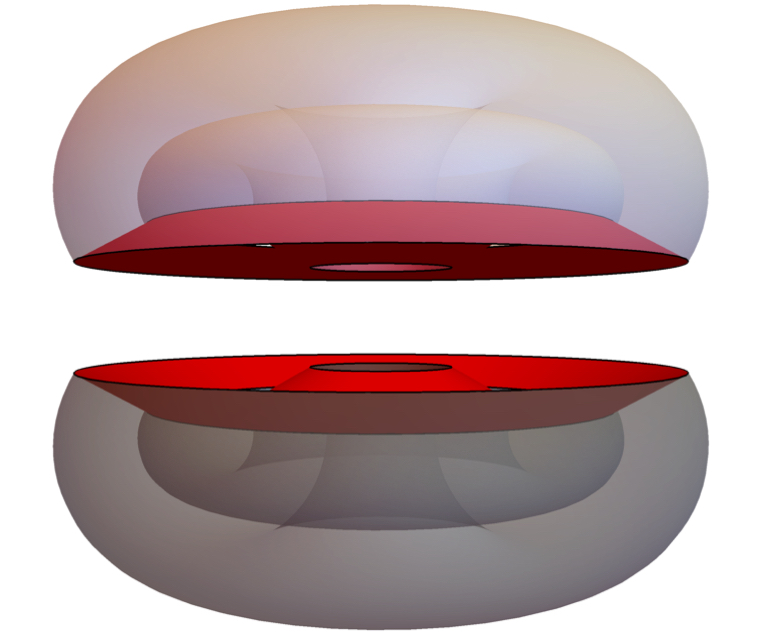}\end{minipage} }
\subfloat[The 3-manifolds are then glued together along the boundary annuli.]
 {\begin{minipage}{.245\textwidth}\vspace{-90pt}\centering\includegraphics[width=.9\textwidth]{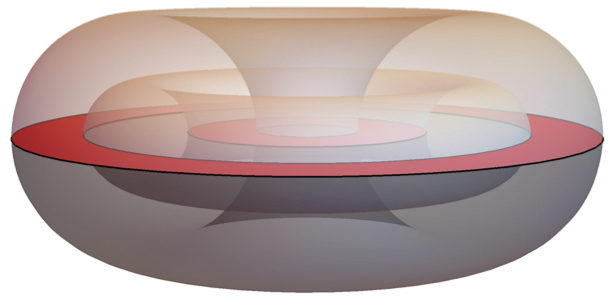}\end{minipage} }
\subfloat[In the $S^3$ picture, this can be algebraically realized using band surgery.]
 {\begin{minipage}{.245\textwidth}\vspace{-90pt}\centering\includegraphics[width=0.95\textwidth]{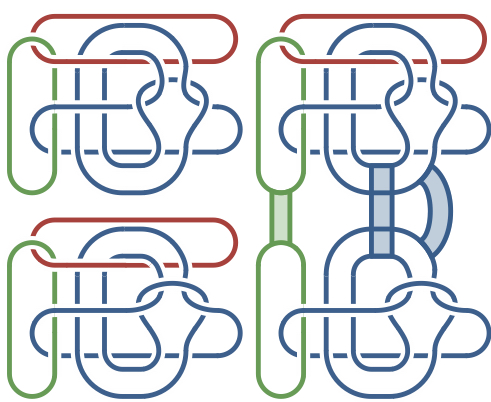}\end{minipage}}
 \subfloat[Garter stitch from an longitudinal annulus sum.]
{\begin{minipage}{0.245\textwidth}\vspace{-90pt}\centering\includegraphics[width=0.75\textwidth]{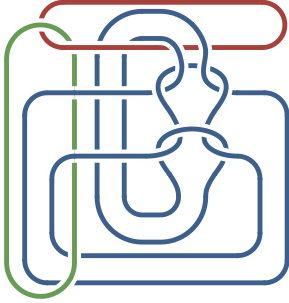} \end{minipage}}

\caption{Annulus sum on the 3-manifold knot (or link) components defines the procedure for combining knit and purl stitches into more complicated 2-periodic textiles.}
 \label{fig:annulus}
 \vspace{-5pt}
\end{figure}

Each of the fabrics in Figure \ref{fig:knit_purl} are 2-periodic and can be made by combining knit and purl stitches either laterally -- as in $1\times1$ ribbing shown in Figures \ref{fig:horiz_join}, vertically -- as in garter, or both -- as in seed. Stockinette and reverse stockinette are represented by knots in $\mathbb{T}^2\times I$ (or links in $S^3$). We would like to create a surgery on these knots (or links) that combines knit and purl stitches to create other 2-periodic textiles. We construct a method for combining stitches using an \emph{annulus sum}. Figure \ref{fig:annulus}a-d illustrates an longitudinal annulus sum, and Figure \ref{fig:annulus}e-h demonstrate the meridional annulus sum. Consider two knit knots $\mathcal{K}_1$ and $\mathcal{K}_2$. These can either be viewed as two disjoint 3-manifolds $\mathbb{T}^2\times I - \mathcal{K}_1$ and $\mathbb{T}^2\times I - \mathcal{K}_2$ or as the 3-manifold created by the complement of two disjoint auxiliary links $\mathcal{L}_1$ and $\mathcal{L}_2$ in $S^3$. The annulus sum is a process to join the disjoint manifolds (or links in $S^3$) into a single knit knot, either both along their meridians or their longitudes.

Adding stitches horizontally involves cutting two tori along their meridians in the $\mathbb{T}^2\times I$ picture, or along the annulus bounded by the green component of the Hopf link in Figure \ref{fig:T2xI}g in the $S^3$ picture, see Figure \ref{fig:annulus}a. In the $\mathbb{T}^2\times I$, cutting each 3-manifold along along its meridian leaves two boundary annuli, punctured by the knit knot. In Figure \ref{fig:annulus}b) each pair of annuli are glued together and the knit knot boundaries are identified. In the $S^3$ picture, the link complements are split along disks that span the meridional (green) component of their Hopf links. These disks are then glued together, identifying the punctures made by the knitted link components. This is equivalent to doing a pair of band surgeries on the links, shown in Figure \ref{fig:annulus}c. The resulting knitted component of the link still has pairwise linking number one with the meridional (green) link component and is trivially linked with the longitudinal (red) component. Therefore, the knitted link component is still has trivial homology around the meridian. Figure \ref{fig:annulus}d shows simple example of this is joining a knit link with a purl link along their meridians to create $1\times1$ ribbing.

Likewise, stitches can also be joined vertically. This process involves joining two thickened tori by cutting along their longitudes, as shown in Figure \ref{fig:annulus}e. The resulting annular boundary components are joined together with the knitted (blue) link punctures identified (as in Figure \ref{fig:annulus}f). In the $S^3$ picture, this involves cutting the 3-manifold along the disks spanning the longitudinal (red) component of the Hopf link and gluing the manifold together along those boundaries (see Figure \ref{fig:annulus}g). This is equivalent to performing three band surgeries on the knit links. The vertical annulus sum adds a component to the link. This component corresponds to another knitted knot. Each of these components link with the meridional (green) component of the Hopf link and they are pairwise unlinked with each other and with the longitudinal (red) component of the Hopf link. Garter fabric can be created by joining a knit link with a purl link along their longitudes, as seen in Figure \ref{fig:annulus}h.

Meridional and longitudinal annulus sums commute. For instance, the checkerboard lattice seen in seed fabric in Figure \ref{fig:knit_purl}e can be created by first creating two tori longitudinally with garter links in them and joining them with a meridional annulus sum. The result is homeomorphic to the link generated by first creating two tori meridionally with $1\times1$ ribbed knots in them and then joining them together with a longitudinal annulus sum.


\section*{Some stitch patterns cannot be made using the annulus sum}
\begin{wrapfigure}[23]{r}{0.45\textwidth}
\vspace{-20pt}
\subfloat[A (left) knot diagram for (right) basketweave fabric shows pairs of stitches that have been swapped, left leaning on odd rows and right leaning on even ones.]{\includegraphics[width=0.44\textwidth]{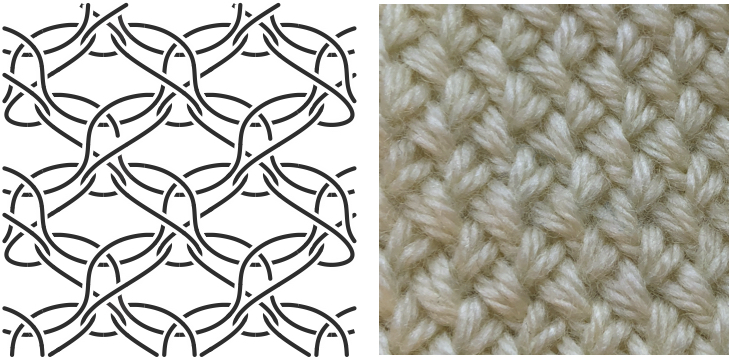} }\\

\vspace{-5pt}\subfloat[This idea can be extended to create braided cables often seen in aran sweaters.]{\includegraphics[width=0.44\textwidth]{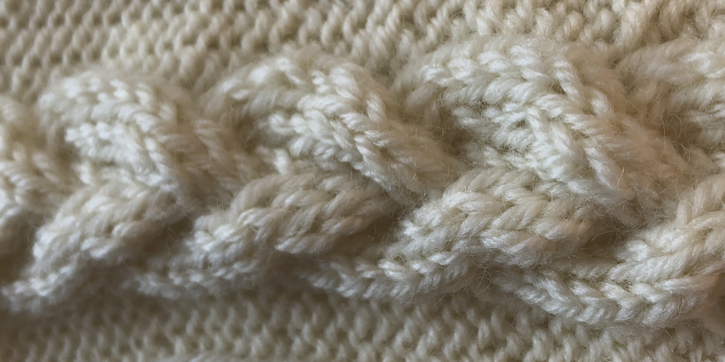} }
\vspace{-5pt}
\caption{Some knitable stitches  cannot be made using the annulus sum.} 
 \label{fig:cables}
 \end{wrapfigure}
There are other topologically allowed knitted stitches that respect the 2-periodic nature of textiles. These occur when the order of stitches within a given row is changed. In knitting, this is known as \emph{cabling}. When stitches are moved, they can create either left leaning or right leaning crossings, when viewed with the wale direction vertically aligned. This creates an algebra of the rows that is analogous to the braid group of $n$ strands. The generators of the braid group are denoted $\sigma$, where $\sigma_i$ acts on strands $i$ and $i+1$ to cross strand $i$ over $i+1$; likewise, $\sigma_i^{-1}$ crosses strand $i+1$ over $i$. For instance, the basketweave pattern in Figure \ref{fig:cables}a is generated on even rows by $\sigma_1\sigma_3\sigma_5...\sigma_n$ and on odd rows by $\sigma^{-1}_2 \sigma^{-1}_4\sigma^{-1}_6...\sigma^{-1}_{n-1}$. The knotted topology of the knitted stitches also changes the algebraic structure of the braid group, such that for subsequent rows, it no longer has an inverse $\sigma_i\sigma^{-1}_i\neq1$. This implies that the structure of the knitted equivalent of the braid group is not a group but a monoid. This is the set of transpositions of a string of $n$ elements. Within a single row, any action of the braid group is valid until they are locked into place by the subsequent row of stitches.

Cabling is a manipulation of stitches that can't be created by using the annulus sum process shown in Figure \ref{fig:annulus}. {We will construct a type of surgery on the manifold that allows us to create transpositions between elements.} It is necessary to keep in mind that, as with braids transpositions have a sense of orientation, either element $i$ passes over $i+1$ or vice versa. {We will incorporate these transformations into the connected sum algebra} we have created for addition of different stitches into a period fabric. A single transposition, as in Figure \ref{fig:cables}a, involves interchanging two stitches. However, in more complicated cables, e.g. the braided cable in Figure \ref{fig:cables}b, two groups of consecutive stitches are interchanged, but this does not need to happen pairwise.

\begin{figure}[t!]
\centering
\subfloat[An $m\times n$ unknit begins with $n$ disjoint, unlinked longitudinal circles and $m$ disjoint circles with trivial homology.]
 {\includegraphics[width=0.245\textwidth]{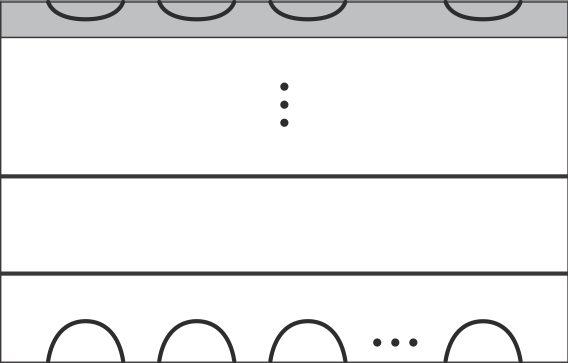} }
\subfloat[Ambient isotopy between bights of the disjoint circles and the longitudinal circles.]
 {\includegraphics[width=0.245\textwidth]{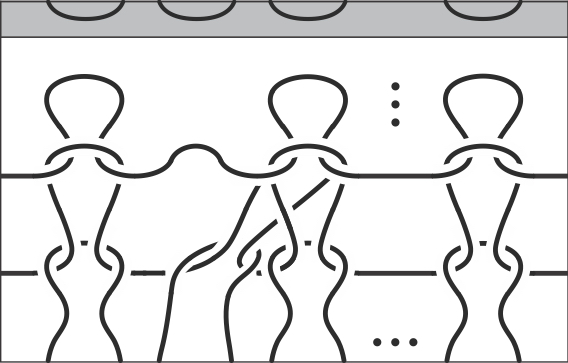} }
\subfloat[Band surgery joins the $m$ disjoint circles to one of the $n$ longitudinal circles.]
 {\includegraphics[width=0.245\textwidth]{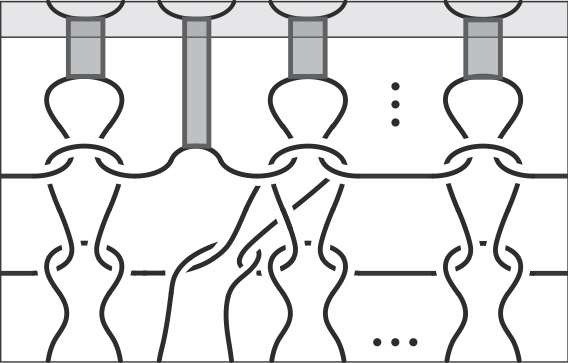} }
 \subfloat[The the $m\times n$ {\rm swatch} in $\mathbb{T}^2\times I$.]
  {\includegraphics[width=0.245\textwidth]{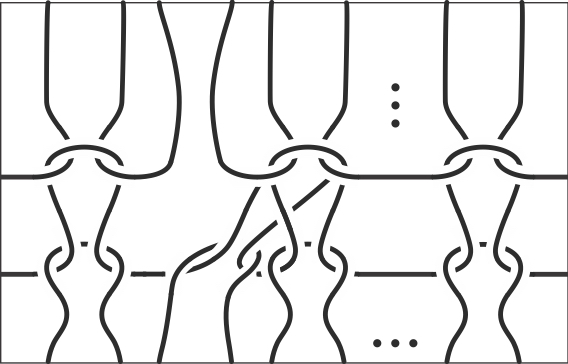} }

\vspace{-10pt}
\caption{Construction of and $m \times n$ {\rm swatch}.}
 \label{fig:swatch_construction}
\end{figure}

Although these more complicated multi-stitch objects cannot be constructed from basic knit and purl elements using annulus sums, they do fit into our framework of links in $\mathbb{T}^2\times I$. This construction, which we call a \emph{swatch} begins with an $n$-stranded \emph{unknit}, made from $n$ disjoint circles along the longitude of the torus and $m$ disjoint circles with trivial homology, see Figure \ref{fig:swatch_construction}a. Figure \ref{fig:swatch_construction}b shows that shows bights of each of the $m$ circles interacting via ambient isotopy with one or more of the $n$ longitudinal strands. These strands are now able to interact with one another via ambient isotopy. Note that this procedure does not change the pairwise linking number of any of the circles. Finally, each of the $m$ circles are joined by band surgery to bights in the last longitudinal strand (Figure \ref{figLswatch_construction}c) to create the final swatch in $\mathbb{T}^2\times I$ (Figure \ref{fig:swatch_construction}d). As the swatches live in $\mathbb{T}^2\times I$, an $k \times n$ swatch and an $l \times n$ swatch can be joined via a meridional annulus sum to create a $(k+l)\times n$ swatch. Likewise, $m\times k$ and $m\times l$ swatches can be joined longitudinally to create an $m\times (k+l)$ swatch.

It is easy to see that all of the objects we have considered thus far fit into this swatch construction. The basic knit and purl are types of $1\times1$ swatch, as is the cow-hitch. $1\times1$ ribbing is a $2\times1$ swatch, while garter is a $1\times2$ swatch. The basketweave structure in Figure \ref{fig:cables}a is a $4\times2$ swatch. This construction shows that all knitted link components are ribbon. However, we can easily show that not all ribbon knots are knitable. For example, we can take the connected sum of a ribbon knot with any of the $n$ longitudinal circles. The resulting knot in $\mathbb{T}^2\times I$ is ribbon, but it is no longer knitable.

\section*{Summary and Conclusions}

Here, we presented a topological framework for 2-periodic knitable structures as knots in $\mathbb{T}^2\times I$ (or as a link in $S^3$). Using meridional and longitudinal annulus sums, we can join different primitive knit elements together to create more complex textiles, including $1\times1$ ribbing, garter and seed fabrics. Knits allow for multiple stitches between rows to interact with each other in non-pairwise ways, thus annulus sums cannot create all possible knits. We define the swatch as a way to construct knitable objects in $\mathbb{T}^2\times I$. Multiple swatches can be joined together using the annulus sum to create more textiles.

\section*{Acknowledgements}

The authors were partially supported by National Science Foundation grant DMR-1847172. The second author was in residence at ICERM in Providence, Rhode Island, during a portion of this work which was supported by National Science Fundation under Grant No. DMS-1439786. We would like to thank sarah-marie belcastro, Michael Dimitriyev, Jen Hom, Jim McCann, Agniva Roy, Saul Schleimer and Henry Segerman for many fruitful conversations.

    
{\setlength{\baselineskip}{13pt} 
\raggedright				

}

\end{document}